\documentclass{pasj01}
\usepackage{natbib, aas_macros}
\draft
\usepackage{lscape}
\usepackage{graphicx}
\Received{$\langle$reception date$\rangle$}
\Accepted{$\langle$acception date$\rangle$}
\Published{$\langle$publication date$\rangle$}
\usepackage{mediabb}

\begin{document}

\title{Suzaku Observations of Low Surface Brightness Cluster Abell 1631}
\author{Yasunori \textsc{Babazaki}\altaffilmark{1,}$^{*}$, Ikuyuki \textsc{Mitsuishi}\altaffilmark{1}, Naomi \textsc{Ota}\altaffilmark{2}, Shin \textsc{Sasaki}\altaffilmark{3}, Hans \textsc{{B{\"o}hringer}}\altaffilmark{4}, Gayoung \textsc{Chon}\altaffilmark{4}, Gabriel W. \textsc{Pratt}\altaffilmark{5,6} and Hironori \textsc{Matsumoto}\altaffilmark{7}} %

\altaffiltext{1}{Department of Physics, Nagoya University, Aichi 464-8602, Japan}
\altaffiltext{2}{Department of Physics, Nara Women's University, Kitauoyanishi-machi, Nara, Nara 630-8506, Japan}
\altaffiltext{3}{Department of Physics, Tokyo Metropolitan University, 1-1 Minami-Osawa, Hachioji, Tokyo 192-0397, Japan}
\altaffiltext{4}{Max-Planck-Institut f{\"u}r extraterrestrische Physik, D-85748 Garching, Germany}
\altaffiltext{5}{IRFU, CEA, Universit{\'e} Paris-Saclay, F-91191 Gif-sur-Yvette, France}
\altaffiltext{6}{Universit{\'e} Paris Diderot, AIM, Sorbonne Paris Cit{\'e}, CEA, CNRS, F-91191 Gif-sur-Yvette, France}
\altaffiltext{7}{Department of Earth and Space Science, Osaka University, Osaka 560-0043, Japan}
\email{y\underline{ }babazaki@u.phys.nagoya-u.ac.jp}


\maketitle

\begin{abstract}
We present analysis results for a nearby galaxy cluster Abell 1631 at $z~=~0.046$ using the X-ray observatory Suzaku. 
This cluster is categorized as a low X-ray surface brightness cluster. 
To study the dynamical state of the cluster, we conduct four-pointed Suzaku observations 
and investigate physical properties of the Mpc-scale hot gas associated with the A1631 cluster for the first time.
Unlike relaxed clusters, the X-ray image shows no strong peak at the center and an irregular morphology. 
We perform spectral analysis and investigate the radial profiles of the gas temperature, density, 
and entropy out to approximately 1.5~Mpc in the east, north, west, and south directions by combining with the XMM-Newton data archive. 
The measured gas density in the central region is relatively low (${\rm a~few} \times~10^{-4}~{\rm cm^{-3}}$) at the given temperature ($\sim2.9~{\rm keV}$) compared with X-ray-selected clusters. 
The entropy profile and value within the central region ($r<0.1~r_{200}$) are found to be flatter and higher ($\gtrsim400~ {\rm keV~cm}^2$). 
The observed bolometric luminosity is approximately three times lower than that expected from the luminosity-temperature relation in previous studies for relaxed clusters.
%
These features are also observed in another low surface brightness cluster, Abell 76. 
The spatial distributions of galaxies and the hot gas appear to be different. 
The X-ray luminosity is relatively lower than that expected from the velocity dispersion.
A post-merger scenario may explain the observed results. 
\end{abstract}

\section{INTRODUCTION}
According to the study of the hierarchical-structure formation, galaxy clusters are the largest collapsed objects in the Universe. 
Galaxy clusters are considered to grow through a complex process that involves the accretion of matter, diffuse inter-galactic medium, 
from filaments and cluster merging events. 
In fact, a variety of cluster morphologies have been observed and are often classified as relaxed and unrelaxed (or merging) clusters.
In the X-ray band, bright clusters with centrally peaked cool-core emission or intense emission due to gas compression via mergers, 
which is predicted by cluster-merger simulations \citep{2001ApJ...561..621R}, usually are easily detected. 
However, faint clusters with rather irregular, diffuse morphologies are difficult to detect and it is no surprise 
that a majority of them may be missed in existing X-ray surveys, and therefore await discovery. 
As we mention below, a population of clusters has an extremely low surface brightness and the evolutionary stage is not yet clarified. 

In the cluster sample detected by the ROSAT All-Sky Survey \citep{2013A&A...555A..30B,2017AJ....153..220B}, 
a small fraction ($\sim5\%$--$10\%$) of galaxy clusters exhibits very low surface brightness profiles and highly diffuse emission. 
These are called ``low surface brightness clusters'' (LSB clusters). In the REXCESS sample, which consists of 33 clusters 
and which is compiled from the southern ROSAT All-Sky cluster survey REFLEX with a morphologically unbiased selection 
in the X-ray luminosity and redshift \citep{2007A&A...469..363B}, three such LSB clusters were identified. 
The LSB clusters tend to show highly irregular morphology with no prominent central core, while a majority of 
known clusters show closely self-similar X-ray surface-brightness profiles \citep{2004A&A...428..757O,2008A&A...487..431C}.

The measurement of entropy distribution of the intracluster medium (ICM) provides a clue to study 
the thermodynamical history of cluster evolution \citep{2005RvMP...77..207V}. 
The three LSB clusters in the above-mentioned cluster sample show relatively higher temperature 
than those of the X-ray selected clusters at a given gas density. In other words, the LSB clusters 
show relatively high central entropy ($K> 200\ {\rm keV\ cm^2}$)\footnote{Defined as $K \equiv kT n_{\rm e}^{-2/3}$, 
where $k$, $T$ and $n_{e}$ are the Boltzmann constant, the temperature, and the electron density, respectively.}.
A possible interpretation of this character is that the clusters are dramatically younger. 

A spatially resolved X-ray spectral analysis in the LSB cluster Abell 76 was conducted by \citet{2013A&A...556A..21O}. 
A76 shows one of the lowest surface brightness values among the ROSAT clusters studied by \citet{1999A&A...348..711N}. 
The entropy within the central region ($r~<~0.2~r_{200}$) is exceptionally high ($\sim400~{\rm keV~cm^2}$). 
This phenomenon is not readily explained by either gravitational heating or preheating. 
The X-ray morphology is clumped and irregular, and the electron density is extremely low ($\sim10^{-3}~{\rm cm^{-3}}$) 
for the observed high temperature ($\sim3.3~{\rm keV}$). 
The diffuse nature of the cluster may suggest that this is the system in formation. 
Currently, the only LSB cluster to be analyzed in detail is A76. 
To clarify the nature of LSB clusters and their thermodynamic evolution, many more clusters should be analyzed.

To further explore the nature of LSB clusters, we focus on Abell 1631 at $z~=~0.046$. 
The cluster has the lowest surface brightness in a cluster sample of ROSAT All-Sky Survey 
and the flux is $\sim 7.8~\times~10^{-12}~{\rm erg~s^{-1}~cm^{-2}}$ 
in the energy range of 0.5--2.4~keV \citep{1996MNRAS.283.1103E}. 
A faint diffuse emission around the cluster was detected by XMM-Newton \citep{2012A&A...545A.140T}; 
however, any detailed spectral analysis of the ICM has not been conducted. 
The rich optical data are also available to investigate the galaxy distribution and 
the velocity dispersion and compare X-ray properties with optical properties to study the cluster dynamics.

The paper is organized as follows: Section 2 presents the observation of A1631 with Suzaku and 
XMM-Newton and the data reduction, and section 3 describes our analysis method and the results. 
In section 4 we discuss the X-ray properties and the optical properties and propose a possible scenario. 
Where necessary, we assume the standard cosmological model with a matter density of ${\rm \Omega}_M~=~0.27$, 
the cosmological constant ${\rm \Omega}_\Lambda~=~0.73$ and the Hubble constant $H_0~=~70~{\rm km~s^{-1}~Mpc^{-1}}$. 
At the cluster redshift $z~=~0.046$, $1\arcmin$ corresponds to 54~kpc. In this paper, we used HEAsoft v6.21
and XSPEC version 12.9 to perform $\chi^2$ fitting and the metal-abundance table of \citet{1989GeCoA..53..197A}. 
Unless otherwise stated, the error ranges show the 90\% confidence level from the center value.
\begin{table*}[htb]
\tbl{Logs of Observations of A1631 with Suzaku and XMM-Newton}{%
\begin{tabular}{llllll}
\hline \noalign{\vskip2pt} 
Target               & Obs ID     & Date                &  RA.     &   Dec.   & Net Exposure\footnotemark[$*$] \\
                     &            &                     & [deg]    &  [deg]   &  [ks]     \\ \hline
Suzaku/XIS           &            &                     &          &          &       \\              
A1631 C             & 807004010 & 2013-01-12 11:47:03 & 193.2413 & -15.3446 & 23.8 \\
A1631 E             & 807005010 & 2013-01-13 00:16:01 & 193.4827 & -15.4429 & 20.5 \\
A1631 W             & 807006010 & 2013-01-13 17:22:21 & 193.3425 & -15.1144 & 21.0 \\
A1631 N             & 807003010 & 2013-01-11 15:54:29 & 193.0052 & -15.2480 & 28.0 \\ \hline
XMM-Newton/EPIC      &            &                     &          &          &       \\
NGC 4756             & 0551600101 & 2008-12-23 19:50:30   &  193.15458   &  -15.4742  &  34.6 (MOS1), 37.8 (MOS2), 20.8 (pn) \\
\hline
\end{tabular}\label{tab:obs_summary}
}
\begin{tabnote}
\footnotemark[$*$] Net exposure time after data filtering
\end{tabnote}
\end{table*}

\section{OBSERVATION AND DATA REDUCTION}
\subsection{Suzaku}\label{sec:reductionsuzaku}
Four-pointed observations of A1631 (the center, east, west, and north) were conducted by Suzaku. 
The details of the observations are summarized in table 1. The XIS instruments, which are X-ray sensitive CCD cameras, 
are installed in Suzaku. These consist of four cameras: three front-illuminated (XIS-0, -2, -3) 
and one back-illuminated (XIS-1) CCD cameras \citep{2007PASJ...59S..23K}. 
Since the entire imaging area of the XIS-2 was lost due to micro-meteorite hits, 
no useful data have been obtained with XIS-2 since 2006 November 9. 
The XIS-0, -1 and -3 CCD cameras were operated in normal mode with the running of 
space charge injection \citep{2009PASJ...61S...9U}.

Event files were created with Suzaku pipeline processing. 
The data were analyzed with CALDB (v2016-06-07 for XIS and v2011-06-30 for XRT) 
which are calibration files for XIS and the X-ray telescopes (XRT: \citet{2007PASJ...59S...9S}). 
The XIS data were filtered by using the criteria as follows: Earth elevation angle $>5^\circ$, 
day-Earth elevation angle $>20^\circ$, and time removal during the South Atlantic Anomaly. 
We applied an additional correction to remove flickering pixels\footnote{https://heasarc.gsfc.nasa.gov/docs/suzaku/analysis/xisnxbnew.html}. 
Point sources detected in the XMM-Newton observation (see, section 2.2) were removed.

In recent years, Suzaku observations have been increasingly affected by the ${\rm O~_{I}}$ contamination 
at 0.525 keV \citep{2014PASJ...66L...3S} caused by the fluorescence of solar X-rays with neutral oxygen in Earth's atmosphere. 
The ${\rm O~_{I}}$ contamination sometimes affects the spectral analysis in estimating X-ray background components. 
Although this effect can be minimized by using events taken during time intervals in the larger elevation angle 
from the bright Earth limb (the DYE\_ELV parameter), we modeled the ${\rm O~_{I}}$ line with a Gaussian instead of 
conducting the DYE\_ELV filtering in the spectral analysis not to reduce the exposure time.
We conducted the spectral analysis using DYE\_ELV filtering and confirmed that the resulting parameters are consistent 
with each other within the statistical errors.

\subsection{XMM-Newton}
We obtained the XMM-Newton/EPIC data for A1631 available from the XMM-Newton Science Archive. 
The XMM-Newton data mainly cover the south region of A1631. 
The EPIC observation data were analyzed with ESAS (Extended Source Analysis Software) \citep{2008A&A...478..615S}, 
packaged in SAS version 16.0.0. 

The EPIC data were processed and screened in the standard way with SAS pipeline software tools epchain, emchain, 
and the ESAS tools pn-filter and mos-filter. The data were filtered by removing time intervals of 
high background due to soft proton flares whose rates were out of the 2$\sigma$ range of the rate distribution. 
The observation identification and the flare-filtered exposure time are listed in table 1. 
By using cheese tool, point sources are identified in each detector and removed from our analysis.

\section{ANALYSIS and RESULTS}\label{sec:analysisandresults}
\subsection{Images}
\begin{figure*}[htb]
  \begin{center}
    \includegraphics[width=170mm,angle=0]{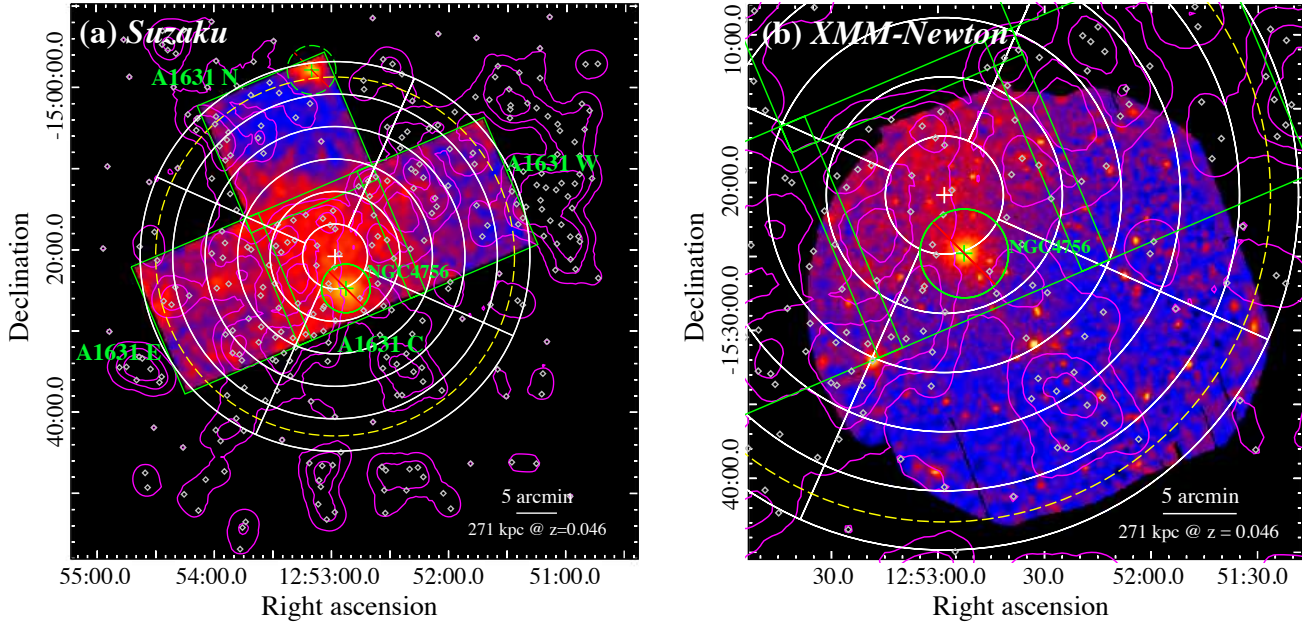}
  \end{center}
  \caption{Suzaku and XMM-Newton X-ray images of A1631 in the energy band 0.5--5 keV. 
The extraction regions used in section 3.2 are shown by white rings with the center of the X-ray peak position indicated by a white cross, 
which is determined by using the XMM-Newton image. 
We excluded from our analysis the region of a foreground elliptical galaxy NGC 4756 ($z~=~0.013599$) 
and shown as a green circle with the center indicated by a green cross. 
The region shown as a dashed green circle corresponds to an F-type star HIP 62872, which is also excluded. 
The estimated virial radius derived from the gas temperature is shown by the yellow dashed circle. 
The green box shows the Suzaku field of view in each observation. 
The 288 spectroscopically confirmed member galaxies are indicated by gray diamonds. 
The magenta contour shows a smoothed member galaxy distribution. (a) Suzaku XIS-1 mosaic X-ray image. 
The NXB image is subtracted. 
The image is corrected for exposure map and the vignetting effect and is smoothed by a Gaussian function with $\sigma~=~25''$. 
(b) Combined XMM-Newton EPIC QPB-subtracted, exposure-corrected, adaptively-smoothed image.
}
\label{fig:img}
\end{figure*}

Figure \ref{fig:img} (a) shows a Suzaku/XIS1 image of A1631 in the 0.5--5 keV band. 
The image was corrected for exposure map and vignetting and was smoothed by a Gaussian function with $\sigma~=~25''$ 
after subtracting a non X-ray background (NXB) image. 
The exposure map and NXB image were generated by using xissim \citep{2007PASJ...59S.113I} 
and xisnxbgen tools \citep{2008PASJ...60S..11T}, respectively. 
Figure \ref{fig:img} (b) shows a combined XMM-Newton/EPIC image in an energy band of 0.5--5.0 keV. 
The quiescent particle background (QPB) image, which was produced by pn\_back and mos\_back tools\citep{2008A&A...478..575K}, 
was subtracted in each image. 
The adapt\_2000 tool was used to create exposure-corrected, QPB-subtracted, adaptively smoothed mosaic images.

From the images, we find clear X-ray diffuse emission from A1631 on the north of NGC4756.
The spatial structure of the X-ray emission in A1631 is clearly detected. 
The cluster's X-ray peak position, indicated by a white cross in figure \ref{fig:img}, 
is determined from the XMM-Newton image since the Suzaku image is affected by contamination 
from the foreground galaxy NGC 4756 due to the large ($\sim 2^\prime$) Suzaku's point spread function (PSF). 
We find that the X-ray morphology shows no strong peaked feature and irregular morphology.

\subsection{Spectra}\label{sec:spectralfit}
To investigate the spatial distribution of the spectral properties, the spectra were extracted from 
the six concentric rings azimuthally (Region 1: $0^\prime < r < 4^\prime$, Region 2: $4^\prime< r < 8^\prime$, 
Region 3: $8^\prime<r<12^\prime$, Region 4: $12^\prime< r< 16^\prime$, 
Region 5: $16^\prime<r<20^\prime$ and Region 6: $20^\prime<r<24^\prime$) as shown in figure \ref{fig:img} except for 
Region 1 because of the low statistics. 
The Suzaku and XMM-newton observation data were used for the spectral analysis in three directions (east, west, and north) 
and the south direction, respectively.
We excluded the region of a foreground galaxy NGC 4756 detected by \citet{2012A&A...545A.140T} 
shown in figure \ref{fig:img} from our spectral analysis. 
An F-type star HIP 62872 is located around the X-ray peak in the corner of the field of view in the north.
The Suzaku spectrum is described by an optically-thin thermal plasma model with $kT\sim0.6~{\rm keV}$ 
and the 0.5--2.4 keV luminosity of $\sim8~\times~10^{28}~{\rm erg~s^{-1}}$ assuming the distance to the star, 65.8 pc, \citep{1997A&A...323L..49P}, 
and we concluded the X-ray emission is originated from the F type star \citep{1999A&A...348..161P}.
The region was also excluded. 

For each annulus in the east, north, and west, the observed 0.5--8 keV Suzaku spectra of 
three sensors (XIS-0, XIS-1, and XIS-3) were simultaneously fitted.
Here the redistribution matrix files (RMFs) were generated by xisrmfgen \citep{2007PASJ...59S.113I}. 
The effective area ancillary response files (ARFs) were calculated by xissimarfgen \citep{2007PASJ...59S.113I}. 
As input images, we used flat-field emission models over the spectra extraction regions. 
The NXB spectra were subtracted by using xisnxbgen tool \citep{2008PASJ...60S..11T}. 

The 0.5--11.0 keV XMM-Newton spectra in the south direction and response files were prepared by the XMM-ESAS tools mos-spectra and pn-spectra. 
The QPB spectra were estimated by pn\_back and mos\_back tools and subtracted from the observed spectra. 
The observed XMM-Newton spectra of MOS1, MOS2 and pn were simultaneously fitted.
We note that only spectra of MOS2 and pn were used in the spectral analysis for Region 1 since the region is located on the lost CCD chip in MOS1.
Instrumental fluorescence lines and the residual soft-proton contamination were modeled with 
Gaussian components and a power-law component, respectively.

To estimate the X-ray background emission, the ROSAT All-Sky Survey spectrum derived from 
the HEASARC X-ray Background Tool\footnote{http://heasarc.gsfc.nasa.gov} was used because 
it is difficult to estimate the emission in the field of view due to the widely distributed ICM.
As X-ray background emissions, the cosmic X-ray background (CXB) and the Galactic foreground emissions arising from the local hot bubble (LHB), 
and the Galactic halo (GH) were taken into account. 
For our analysis, the RASS spectrum was extracted from a 1-2 degree annulus surrounding A1631. 
We modeled the X-ray background components with the formula `` apec$_{\rm LHB}$ + phabs * (apec$_{\rm GH}$ + power\-law$_{\rm CXB})$'', 
where APEC and PHABS show an optically thin thermal plasma in collisional ionization equilibrium \citep{2001ApJ...556L..91S} 
and a photoelectric absorption model, respectively. The temperature of apec$_{\rm LHB}$ and the Galactic column density $N_{\rm H}$ of PHABS 
based on the HI maps \citep{2005A&A...440..775K} are fixed to 0.1 keV and $3.4~\times~10^{20}~{\rm cm}^{-2}$, respectively. 
The index of power\-law$_{\rm CXB}$ was fixed at 1.4. We confirmed that the RASS spectrum 
is described with the spectral model and the CXB intensity is consistent with the typical value \citep{2002PASJ...54..327K}. 
Thus, we concluded the RASS spectrum is suitable to estimate the X-ray background emission.
To take into account the statistical uncertainty of the X-ray background spectrum, the observed spectra and the RASS spectrum were simultaneously fitted.
We confirmed that resulting parameters do not change significantly within the statistical errors even though the 30 \% uncertainty is taken into account to explain the temporal and spatial fluctuations in the X-ray background components. 
\begin{figure}[htb]
\begin{center}
\hspace{-0.4cm}
    \includegraphics[width=85mm,angle=0]{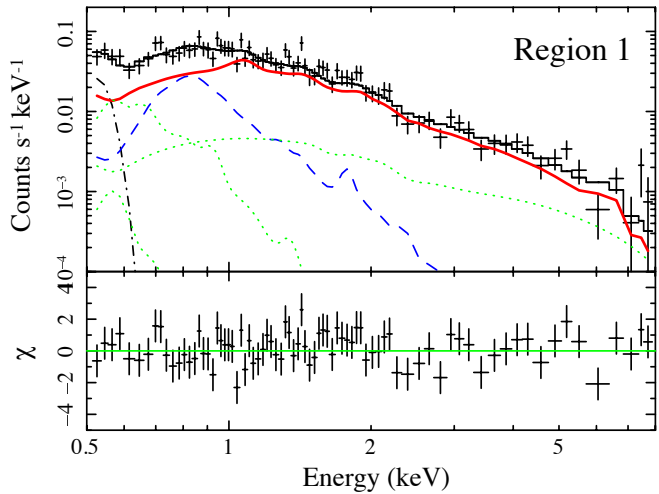}
  \end{center}
\vspace{-.2cm}
\caption{NXB-subtracted XIS-1 spectrum in Region 1 (for definition of regions, see figure \ref{fig:img}) with the best-fit 2T models. 
The (red) bold and (blue) dashed lines are the higher and lower temperature components, respectively. 
The (green) dotted lines show X-ray background components, which are estimated by combining with the RASS spectrum. 
The (black) dash-dotted line at 0.5 keV shows the ${\rm O~_I}$ fluorescent line originating from solar X-rays (see section \ref{sec:reductionsuzaku}).
For simplicity, the XIS-0 and -3 spectra are not shown in this figure, although they were analyzed simultaneously with the XIS-1 spectrum.
 }
\label{fig:spectrum}
\end{figure}

Each observed spectrum was fitted with an absorbed 1T model ``phabs * apec$_{\rm ICM}$'', first.  
The metal abundance was fixed to be the typical value $Z~=~0.3~{\rm solar}$ for nearby clusters 
because the abundance was not well constrained. Some spectra show residuals around $\sim 0.7~{\rm keV}$, 
corresponding to the Fe-L lines, and require an additional plasma model.
The metal abundance and the redshift of the second APEC model were also fixed to ${\rm 0.3~solar}$ and $z ~= ~0.046$, respectively. 
We note that the assumption of the redshift does not affect the spectral-fitting results. 
For Region 1, Region 2-3 south, Region 3 east, Region 4 north and Region 6 east, 
the 2T model was preferred (significance $>3~\sigma$ in an F-test, see table 2). 
An example of a spectrum with the 2T model is shown in figure \ref{fig:spectrum}.
The temperature and the absorption-corrected 0.5--8.0 keV intensity of the two components are summarized in table 2. 
We assessed the systematic errors on the intensity between the detectors, which are $\sim$20\% in both XIS and EPIC instruments 
and are included in the errors of table 2.
The parameters in Region 1 obtained using Suzaku spectra and XMM-Newton spectra are consistent within the errors.

\begin{table*}[ht]
\tbl{Best-fit parameters obtained from the projected analysis}{%
\hspace{0cm}
\begin{tabular}{llcccccc}
    \hline \noalign{\vskip2pt}
 Direction\footnotemark[$*$] &  Parameter\footnotemark[$\dagger$]\footnotemark[$\ddagger$] &  Region 1\footnotemark[$\S$] &  Region 2  & Region 3 & Region 4  & Region 5 & Region 6 \\
           &                    &  $0^\prime < r < 4^\prime$  & $4^\prime < r < 8^\prime$ & $8^\prime < r < 12^\prime$ & $12^\prime < r < 16^\prime$ & $16^\prime < r < 20^\prime$ & $20^\prime < r < 24^\prime$ \\ \hline
E         & $kT_h$ [keV] &  $3.44_{-0.29}^{+0.45}$  & $2.77_{-0.28}^{+0.36}$ & $2.67_{-0.37}^{+0.49}$ &$3.21_{-0.36}^{+0.51}$ &$2.42_{-0.33}^{+0.57}$ & $2.31_{-0.35}^{+0.68}$\\
           & $I_{h}$\footnotemark[$\|$]  [$10^{-7}~{\rm erg~cm^{-2}~s^{-1}~str^{-1}}$] & $3.45\pm0.69$  & $2.82\pm0.56$ & $1.63\pm0.32$  & $1.95\pm0.39$ & $1.01\pm0.20$ & $1.03_{-0.17}^{+0.22}$ \\
           & $kT_l$ [keV] &  $0.75_{-0.16}^{+0.10}$   & -- & $0.88_{-0.12}^{+0.19}$  & --  & --  & $0.77_{-0.11}^{+0.13}$ \\
           & $I_l$\footnotemark[$\|$]  [$10^{-7}~{\rm erg~cm^{-2}~s^{-1}~str^{-1}}$]  &  $0.52\pm0.1$  & -- & $0.28_{-0.12}^{+0.14}$ & -- & -- &$0.42_{-0.17}^{+0.13}$  \\
           & $\chi^2/d.o.f.$ & 203/187 & 114/115 & 133/131 & 184/178 & 142/145 & 84/95 \\ 
           & $kT_1$ [keV]  &$3.19_{-0.30}^{+0.25}$ &$2.72_{-0.25}^{+0.42}$ &$2.35_{-0.24}^{+0.34}$ &$3.22_{-0.35}^{+0.42}$ &$2.50_{-0.34}^{+0.45}$ &$2.07_{-0.26}^{+0.41}$ \\
           & ${\rm Norm}_1^b$ [cm$^{-5}$ str$^{-1}$]  &$347.8_{-25.0}^{+22.0}$ &$245.6_{-21.3}^{+25.6}$ &$185.6_{-16.1}^{+20.4}$ &$162.3_{-13.2}^{+13.3}$ &$97.1_{-11.2}^{+10.9}$ &$138.5_{-19.5}^{+19.9}$ \\
& $chi^2/d.o.f$ &1.19(225.3/189) &0.97(111.5/115) &1.10(146.8/133) &1.05(186.3/178) &0.98(141.7/145) &1.03(99.9/97) \\\cline{2-8}
N          & $kT_h$ [keV] &   &$2.95_{-0.31}^{+0.31}$ &$3.31_{-0.48}^{+0.44}$ & $3.80_{-0.91}^{+1.63}$ & $1.72_{-0.26}^{+0.36}$ &$1.22_{-0.29}^{+0.54}$ \\
           &$I_h$\footnotemark[$\|$]  [$10^{-7}~{\rm erg~cm^{-2}~s^{-1}~str^{-1}}$] & & $3.55\pm0.71$ & $1.98^{+0.17}_{-0.22}$ & $0.69_{-0.11}^{+0.15}$  & $0.30^{+0.06}_{-0.08}$ & $0.29^{+0.12}_{-0.10}$ \\
           & $kT_l$ [keV] & $\uparrow$  & -- & -- & $0.77_{-0.16}^{+0.24}$ & -- & -- \\
           & $I_l$ \footnotemark[$\|$] [$10^{-7}~{\rm erg~cm^{-2}~s^{-1}~str^{-1}}$] & & -- & -- & $0.19_{-0.09}^{+0.08} $ & -- & -- \\
           & $\chi^2/d.o.f.$ &   &116/129 & 175/166 & 179/189 & 149/147  & 127/109  \\ 
           & $kT_1$ [keV]  &$2.90_{-0.26}^{+0.38}$ &$3.21_{-0.43}^{+0.47}$ &$3.05_{-0.56}^{+0.78}$ &$1.70_{-0.31}^{+0.37}$ &$0.94_{-0.11}^{+0.08}$ \\
           & ${\rm Norm}_1^b$ [cm$^{-5}$ str$^{-1}$]  &$303.3_{-24.2}^{+27.0}$ &$156.8_{-14.3}^{+8.8}$ &$70.4_{-8.5}^{+9.7}$ &$29.7_{-8.9}^{+8.7}$ &$46.4_{-9.4}^{+5.9}$ \\
           & $chi^2/d.o.f$ &0.89(114.8/129) &1.08(178.9/166) &0.98(187.0/191) &1.04(153.3/147) &1.18(128.2/109) \\\cline{2-8}
W          & $kT_h$ [keV] &  &$3.14_{-0.42}^{+0.45}$ &$2.68_{-0.38}^{+0.49}$ &$2.83_{-0.62}^{+0.56}$ &$2.58_{-0.46}^{+1.02}$ &$2.55_{-0.64}^{+1.62}$ \\
           &$I_h$\footnotemark[$\|$]  [$10^{-7}~{\rm erg~cm^{-2}~s^{-1}~str^{-1}}$] & & $2.37\pm0.47$ & $1.46\pm0.29$ & $1.03\pm0.21$ & $0.75^{+0.17}_{-0.11}$ & $0.59^{+0.09}_{-0.11}$ \\
           & $kT_l$ [keV] &  $\uparrow$ & -- & -- & -- & -- & --  \\
           & $I_l$ \footnotemark[$\|$] [$10^{-7}~{\rm erg~cm^{-2}~s^{-1}~str^{-1}}$] & & -- & -- & -- & -- & -- \\
           & $\chi^2/d.o.f.$ &  & 113/104 & 119/102 & 146/137 & 140/127 & 97/94  \\
           & $kT_1$ [keV]  &$3.18_{-0.48}^{+0.53}$ &$2.72_{-0.49}^{+0.57}$ &$2.58_{-0.40}^{+0.76}$ &$2.71_{-0.49}^{+0.94}$ &$2.67_{-0.66}^{+1.44}$ \\
           & ${\rm Norm}_1^b$ [cm$^{-5}$ str$^{-1}$]  &$199.7_{-22.3}^{+21.9}$ &$131.0_{-20.1}^{+17.7}$ &$84.6_{-10.8}^{+14.7}$ &$70.1_{-10.3}^{+10.1}$ &$55.5_{-10.5}^{+10.5}$ \\
           & $chi^2/d.o.f$ &1.06(109.9/104) &1.13(115.7/102) &1.03(141.8/137) &1.10(140.3/127) &1.06(99.6/94) \\ \cline{2-8}
S          & $kT_h$ [keV]                                      & $2.57_{-0.18}^{+0.61}$ & $2.11_{-0.44}^{+0.78}$ & $1.66_{-0.21}^{+0.74}$ &  --  &   --   &   --   \\
           & $I_h$ \footnotemark[$\|$] [$10^{-7}~{\rm erg~cm^{-2}~s^{-1}~str^{-1}}$] & $2.37\pm0.47$ & $0.65\pm0.17$ & $0.40_{-0.08}^{+0.10}$ & -- & -- & -- \\
           & $kT_l$ [keV]                                      & $1.03_{-0.18}^{+0.33}$ & $0.92_{-0.09}^{+0.09}$ & $0.84_{-0.07}^{+0.11}$ & $0.95_{-0.06}^{+0.06}$  & $0.67_{-0.10}^{+0.15}$ & $0.65_{-0.24}^{+0.22}$  \\
           & $I_l$ \footnotemark[$\|$] [$10^{-7}~{\rm erg~cm^{-2}~s^{-1}~str^{-1}}$] & $0.20_{-0.12}^{+0.32}$ & $0.36\pm0.11$  & $0.26\pm0.05$ & $0.27\pm0.05$ & $0.14_{-0.05}^{+0.06}$ & $0.10_{-0.05}^{+0.06}$ \\
           & $\chi^2/d.o.f.$                                   & 514/460 &  402/390  & 584/496 & 582/513 &  560/479  &  421/332  \\
    \hline
  \end{tabular}\label{tab:projection}
 }
  \begin{tabnote}
    \footnotemark[$*$] The parameters were obtained by using the following data; the Suzaku data for the east, north, and west directions and the XMM-Newton data for the south direction.\\
    \footnotemark[$\dagger$] If the additional thermal model is needed with significance $> 3 \sigma$ in an F test, we adopt the 2T model.\\
    \footnotemark[$\ddagger$] Subscripts $h$ and $l$ show high and low-temperature components.\\
　　\footnotemark[$\S$] Region 1 was common to the four directions. See the innermost circle in figure \ref{fig:img}.\\ 
    \footnotemark[$\|$] Intensity in the energy band of 0.5--8.0 keV.
  \end{tabnote}
\end{table*}

The high-temperature ($kT\sim2.6~{\rm keV}$) components were significantly detected in the whole regions of the east, north, 
and west directions and only in a central part of the south, whereas the low-temperature ($kT\sim0.7~{\rm keV}$) 
components were detected mostly in the south direction. 
The quite different spatial distribution suggests that each component has a different origin. 
In \citet{2012A&A...545A.140T}, the thermal diffuse X-ray emission with $kT\sim 0.7~{\rm keV}$ in the south-west direction of NGC4756 is detected. 
It has been inferred that the emission is originated in a foreground galaxy group ($z~=~0.01399$) that contains NGC 4756.
On the other hand, there is a possibility that the 0.7 keV excess emission is a part of the Galactic foreground emissions as reported in several previous studies \citep[e.g.,][]{2003ApJ...583...70M,2016PASJ...68S..31S}.  
Thus, we assumed that the low-temperature component detected in Region 1 is uniformly distributed in the whole region, re-conducted spectral analysis, and confirmed that the resulting parameters of the high-temperature components are consistent with those of the original analysis within the statistical errors.
The low-temperature components are much fainter and does not have significant impact on the resulting parameters  
for the high-temperature components.
Because the origin of the low-temperature components is beyond the scope of this paper, we will focus on the high-temperature components and drive the physical properties of the ICM associated with A1631 hereafter. 
Note that all physical parameters described below are those of the high-temperature components.

The mean temperature of the cluster is derived to be 2.9$\pm$0.3 keV using all rings within $1.5~{\rm Mpc}$. 
From the temperature and the relationship given in table 2 of \citet{2005A&A...441..893A}, the virial radius of
 the cluster is estimated to be $r_{200}~=~1.2~{\rm Mpc}$, which corresponds to $\sim22'$ at $z~=~0.046$. 
The absorption-corrected flux within $r~\lesssim~0.7~r_{200}$ in the energy band of 0.5--8 keV is estimated to 
be $7.0~\times~10^{-12}~{\rm erg~s^{-1}~cm^{-2}}$.
The absorption-corrected luminosity in the 0.5--8 keV band and the bolometric luminosity of the cluster are $ 3.5~\times~10^{43}~{\rm erg~s^{-1}}$ 
and $4.5~\times~10^{43}~{\rm erg~s^{-1}}$, respectively. The bolometric luminosity is approximately 1/3 of that expected from 
the observed temperature-luminosity relationships in the X-ray selected clusters \citep[e.g.,][]{2009A&A...498..361P}.

\begin{figure*}[htb]
\begin{center}
    \includegraphics[width=150mm,angle=0]{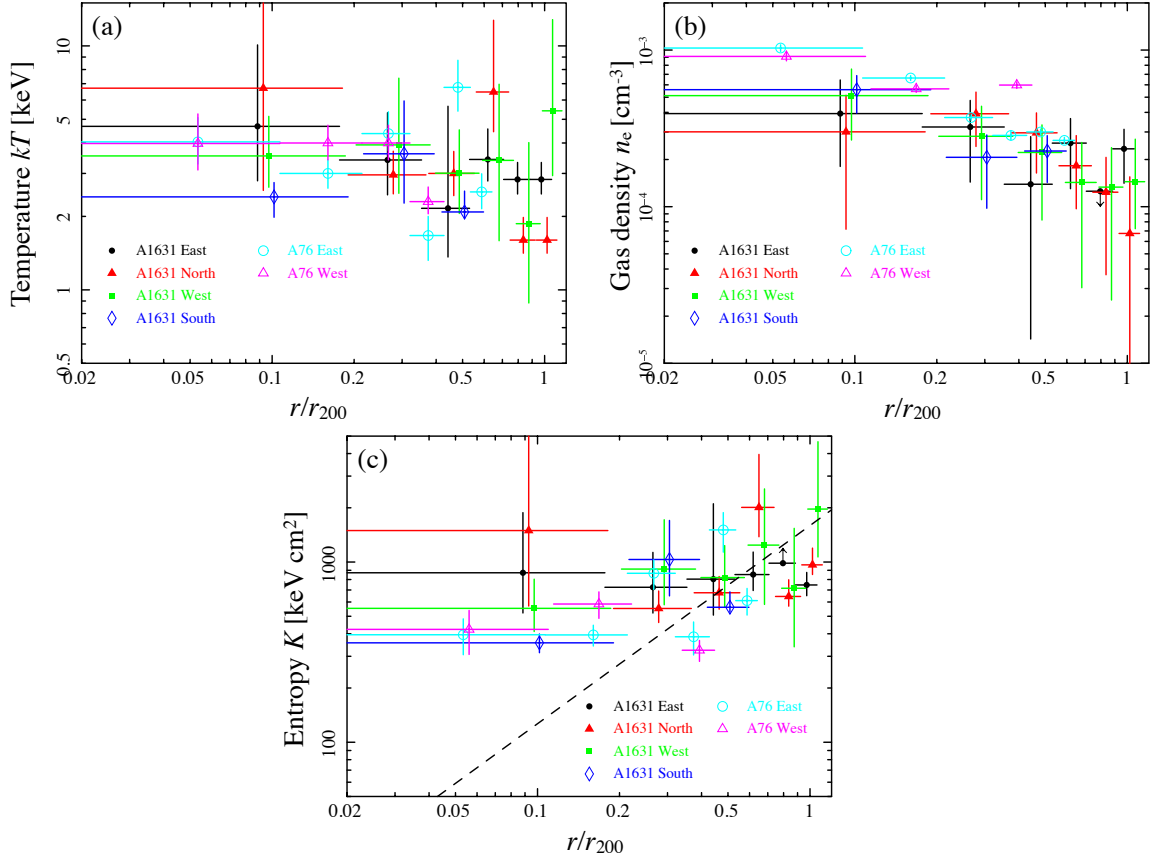}
  \end{center}
\vspace{-0.0cm}
\caption{(a) Temperature, (b) gas density, and (c) entropy profiles for the east (black filled circle), 
north (red filled triangle), west (green filled square), and south (blue empty diamond) derived from the deprojection analysis. 
The horizontal axis is normalized by the virial radius of 1.2 Mpc. 
The markers for the west, north, and south directions are shifted horizontally by 5\%, 10\%, and 15\%, respectively. 
The cyan empty circle and magenta empty triangle marks show another low surface brightness cluster, A76, 
in the east and west directions \citep{2013A&A...556A..21O}. 
In panel (c), the dashed line represents the baseline entropy profile, $K~=~550~{\rm keV~cm^2}~(kT/1~{\rm keV})r^{1.1}$ 
(we assumed the derived $kT~=~2.9~{\rm keV}$), derived in \citet*{2005MNRAS.364..909V}.
 }
\label{fig:properties}
\end{figure*}

\begin{table*}[ht]
\tbl{Temperature, gas density, and entropy obtained from the deprojection analysis of A1631 East, North, West, and South}{%
\hspace{0cm}
\begin{tabular}{llcccccc}
    \hline \noalign{\vskip2pt}
 Direction &  Parameter         &  Region 1 &  Region 2  & Region 3 & Region 4  & Region 5 & Region 6\\
           &                    &  $0^\prime < r < 4^\prime$  & $4^\prime < r < 8^\prime$ & $8^\prime < r < 12^\prime$ & $12^\prime < r < 16^\prime$ & $16^\prime < r < 20^\prime$ & $20^\prime < r < 24^\prime$ \\ \hline
E         & $kT$ [keV]   & $4.68^{+5.41}_{-1.88}$  & $3.41^{+1.91}_{-0.96}$  & $2.16^{+3.50}_{-0.80}$  & $3.42^{+1.14}_{-0.63}$  & --\footnotemark[$*$]  & $2.83^{+0.49}_{-0.36}$ \\
           & $n_{\rm e}$ [$10^{-4}$ cm$^{-3}$]  & $3.93^{+2.52}_{-2.12}$  & $3.23^{+1.54}_{-1.79}$  & $1.39^{+1.41}_{-1.25}$  & $2.55^{+1.11}_{-1.24}$  & $<1.26$  & $2.34^{+0.78}_{-0.93}$ \\
           & $K$ [keV cm$^2$]   & $872^{+1008}_{-351}$  & $724^{+407}_{-203}$  & $804^{+1303}_{-296}$  & $852^{+284}_{-157}$  & $>985$  & $746^{+130}_{-96}$ \\ \cline{2-8}
N          & $kT$ [keV] & $6.71^{+19.01}_{-4.15}$  & $2.96^{+0.74}_{-0.49}$  & $3.00^{+0.69}_{-0.57}$  & $6.47^{+6.26}_{-2.02}$  & --\footnotemark[$\dagger$]  & $1.60^{+0.38}_{-0.19}$ \\ 
           & $n_{\rm e}$ [$10^{-4}$ cm$^{-3}$] & $3.00^{+2.14}_{-2.29}$  & $3.92^{+1.47}_{-1.50}$  & $2.96^{+1.01}_{-1.32}$  & $1.83^{+1.02}_{-0.86}$  & $1.24^{+0.83}_{-0.87}$  & $0.67^{+0.88}_{-0.64}$ \\
           & $K$ [keV cm$^2$]  & $1495^{+4238}_{-926}$  & $553^{+138}_{-91}$  & $675^{+156}_{-128}$  & $2011^{+1947}_{-629}$  & $644^{+153}_{-75}$  & $965^{+230}_{-112}$ \\ \cline{2-8}
W          & $kT$ [keV] & $3.54^{+1.60}_{-0.91}$  & $3.93^{+3.45}_{-1.44}$  & $3.01^{+1.52}_{-0.95}$  & $3.39^{+3.57}_{-1.80}$  & $1.87^{+2.15}_{-0.98}$  & $5.42^{+7.39}_{-2.48}$ \\
           & $n_{\rm e}$ [$10^{-4}$ cm$^{-3}$]  & $5.12^{+2.44}_{-2.46}$  & $2.81^{+1.56}_{-1.71}$  & $2.22^{+1.11}_{-1.40}$  & $1.43^{+1.14}_{-1.12}$  & $1.33^{+1.05}_{-1.08}$  & $1.44^{+1.25}_{-0.72}$\\
           & $K$ [keV cm$^2$] & $553^{+250}_{-142}$  & $914^{+803}_{-335}$  & $820^{+414}_{-259}$  & $1242^{+1305}_{-660}$  & $716^{+823}_{-377}$  & $1968^{+2687}_{-901}$ \\ \cline{2-8}
S         & $kT$ [keV] & $2.41^{+0.35}_{-0.42}$ & $3.61^{+2.34}_{-1.34}$ & $2.09^{+0.45}_{-0.13}$  & -- & -- & -- \\
          & $n_{\rm e}$ [$10^{-4}$ cm$^{-3}$]  & $5.59^{+1.29}_{-1.63}$  & $2.07^{+0.81}_{-1.09}$   & $2.27^{+0.57}_{-0.84}$ & -- & -- & -- \\
          & $K$ [keV cm$^2$] &  $355^{+51}_{-62}$  & $1034^{+669}_{-385}$  & $561^{+122}_{-35}$  & -- & -- & --\\
    \hline
  \end{tabular}\label{tab:deprojection}
 }
  \begin{tabnote}
   \footnotemark[$*$] Linked to the value for $20^\prime < r < 24^\prime$ in the east direction.\\
   \footnotemark[$\dagger$] Linked to the value for $20^\prime < r < 24^\prime$ in the north direction.
   
  \end{tabnote}
\end{table*}

\subsection{Three-dimensional radial profiles}
To derive the three-dimensional structure of the temperature, density, and entropy, 
we conducted a deprojection analysis assuming the spherically symmetric gas distribution. 
The analysis was performed by using ``projct'' model in XSPEC. 
In the deprojection analysis, it is difficult to constrain the parameters of the low-temperature components. 
The parameters were limited to the ranges of statistical errors obtained from the projected analysis 
to take into account the uncertainties.
Region 1 was commonly used in this deprojection analysis of each direction.
Since the uncertainty in temperature is large in the Region 5 for the east and the north, 
the temperature of the regions are linked to those of the outer regions.

The derived temperature profiles with the radius normalized by the virial radius are shown in figure \ref{fig:properties} (a). 
In each direction, the east, west, and south, the temperature coincides from the inner radius to the outer radius within the statistical errors.
For the north direction, the profile shows decrease from $\sim7~{\rm keV}$ to $\sim2~{\rm keV}$ with increasing radius, except for 
the increase around $\sim0.6~r_{200}$.
In figure \ref{fig:properties} (b), the electron density $n_{\rm e}$ is calculated from APEC's 
normalization factor $\int n_{\rm e}n_{\rm H}~{\rm d}V/(4((1 + z) D_A)^2) [10^{-14}~{\rm cm^{-5}}]$, 
where $n_{\rm H}$, $D_{\rm A}$ and $V$ show the proton density ($n_{\rm e}~=~1.2n_{\rm H}$ for the metal abundance of $Z~=~0.3~{\rm solar}$), 
the angular-diameter distance to the source and the volume of the ICMs.
There is no significant azimuthal dependence in the electron density profiles. 
The profiles have radial gradients (from $\sim5~\times~10^{-4}~{\rm cm}^{-3}$ to $\sim1~\times~10^{-4}~{\rm cm}^{-3}$).
The gas entropy profile is evaluated from the temperature and gas density as shown in figure \ref{fig:properties} (c). 
The entropy profile is found to be flat, with an azimuthally averaged central entropy ($r\lesssim0.1~r_{200}$) of $\sim 800~{\rm keV cm^2}$. 
The derived temperature, gas density, and entropy are summarized in table 3.

We assessed the possible systematic errors in the temperature and density as follows: (i) contamination from NGC4756 due to large Suzaku PSF, 
(ii) uncertainty of the metal abundance. 
First, we assessed the systematic error due to (i). 
To estimate the effect of the PSF scattering from NGC4756 with the flux of $5.4~\times~10^{-13}~{\rm erg~s^{-1}~cm^{-2}}$ (0.5--8 keV), 
we used the ray-tracing simulator xissim tool \citep{2007PASJ...59S.113I} to generate Suzaku event files of NGC4756 
by using the method described in \citet{2017ApJ...836..110E}. 
The XMM-Newton/EPIC images and the spectral model with best-fit parameters derived from the analysis of NGC4756 spectra were
used to create simulated event files of each XIS sensor with $5~\times~10^6$ photons. 
The fractions in the 0.5--8 keV band of photons originated from NGC4756 are 3.6\% for Region 1 and 1.2\% for Region 2 east. 
The other regions have the smaller percentage than that of Region 2 east, making them a negligible contribution to the flux from each region. 
To assess the impact of (ii), we fixed the metal abundance to $Z~=~0.5/0.1~{\rm solar}$ and conducted the above-mentioned spectral fit. 
Consequently, the density and temperature of each region are consistent with the results with $Z~=~0.3~{\rm solar}$ 
within the statistical error ranges. 
Thus, we confirmed that the systematic errors do not change our results.

\section{DISCUSSION}\label{sec:discussion}
In this section, to discuss the dynamical state of the cluster, we compare the X-ray properties with 
those of other X-ray selected cluster and examine the optical properties.

\subsection{Temperature, density, and entropy profiles}
\begin{figure}[bht]
\vspace{0cm}
\begin{center}
    \includegraphics[width=90mm,angle=0]{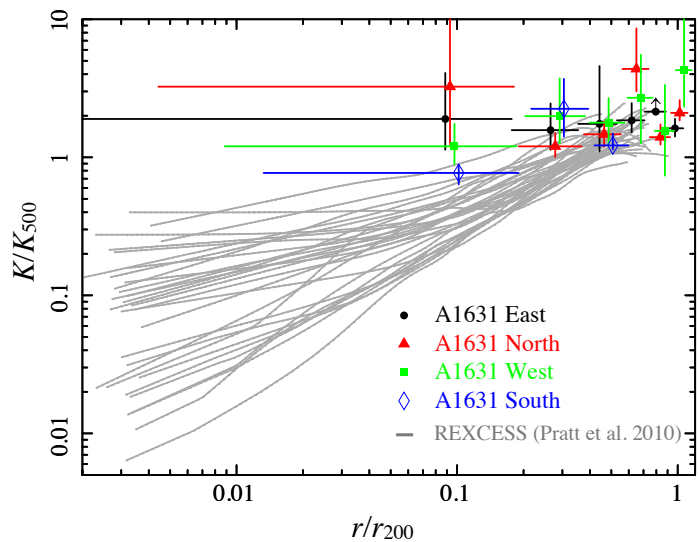}
\end{center}
\caption{Dimesionless entropy profiles of A1631 east (black filled circle), north (red filled triangle), west (green filled square), and south (blue empty diamond) compared to those of the REXCESS sample (gray line; \citet{2010A&A...511A..85P}). The vertical axis is normalized by the $K_{500}$ derived from eq. (3) of \citet{2010A&A...511A..85P}. The horizontal axis is renormalized by $r_{200}$. 
 }
\label{fig:REXCESScomp}
\end{figure}

\citet{2007A&A...461...71P} derived the temperature profiles of the 15 nearby clusters using the XMM-Newton data. 
Comparing our resulting temperature profile (figure \ref{fig:properties} (a)) with theirs, A1631 exhibits the flatter profile. 
The density profiles also show relatively flat, relative to those of REXCESS sample examined in \cite{2008A&A...487..431C}. 
The azimuthally averaged density profile is fitted by the single $\beta$ model with the core radius of $0.14~\pm~0.07~r_{200}$, 
$\beta$ of $0.22~\pm~0.05$, and the central electron density of $(4.9~\pm~0.6)~\times~10^{-4}~{\rm cm^{-3}}$. 
The $\beta$ value is relatively low at a given core radius compared to the other X-ray selected clusters 
\citep[see, figure 18 (a) of ][]{2004A&A...428..757O}.
The central electron density, $\sim5~\times~10^{-4}~{\rm cm^{-3}}$, which is relatively 
lower density than that of other clusters at a given temperature \citep{2012ARA&A..50..353K}. 
The above-mentioned trends are similar to those of A76 \citep{2013A&A...556A..21O}. 
The gas mass within $r_{200}$ is calculated to be $0.34 \times 10^{14} M_\odot$ using the above-described $\beta$ model 
and a total mass within $r_{200}$ of $2.1 \times 10^{14} M_\odot$ estimated from the M-T relation \citep{2005A&A...441..893A}.
Then, the gas-mass fraction is estimated to be $\sim0.16$ and the value is in agreement with universal gas mass fraction of $\sim$0.15 even though the uncertainty in the mass estimation remains considering the dynamical state of the cluster.
Comparing our resulting entropy profile with those of X-ray selected clusters in \citet{2010A&A...511A..85P} (figure \ref{fig:REXCESScomp}), 
we find that A1631 exhibits one of the highest entropies at the center ($r \lesssim 0.1$~$r_{200}$). 

The central entropy ($\gtrsim 400~{\rm keV~cm^2}$) at $0.1~r_{200}$ is at least approximately four times 
larger than that expected from gravitational heating alone ($\sim100~{\rm keV cm^2}$) shown as the dashed line in figure \ref{fig:properties} (c).
Non-gravitational heating processes, such as AGN feedback and preheating, are plausible mechanisms for increasing the entropy.
We assessed the impact of non-gravitational heating processes in the same way with \citet{2013A&A...556A..21O}. 
The excess entropy $\Delta K_{\rm AGN}$ due to AGN feedback are estimated to be $\sim0.4~{\rm keV~cm^2}$ 
by using the empirical $L_{\rm K}$-$\Delta K_{\rm AGN}$ relationship derived by \cite{2010RAA....10.1013W} 
and the K-band luminosity log($L_{\rm K}$/$L_{\rm K, \odot}$,) = 11.1 for PGC 043777 ($z=0.0474$), the brightest member galaxy in the K-band. 
The excess entropy produced by preheating is $\sim140~{\rm keV~cm^2}$ \citep{2013A&A...556A..21O}.
Therefore, it is unlikely that the high entropy $K~\gtrsim~400~{\rm keV~cm^2}$ in A1631 can be explained by the AGN feedback and preheating. 
We note that the possibility remains that the past extraordinary AGN activities injected the high entropy.
\subsection{Comparison between X-ray and optical properties}
The optical information is also useful to discuss the dynamical state of clusters \citep[see, e.g.,][]{2000A&AS..144..187C}.
The cluster was observed as a part of the Omega WIde-field Nearby Galaxy-cluster Survey (OmegaWINGS) 
\citep[see, e.g.,][]{2015A&A...581A..41G,2017A&A...599A..81M}, which is an optical follow-up survey
(photometric and spectroscopic survey) of 76 nearby ($0.04 < z < 0.07$) X-ray selected clusters out to the virial radius 
and the redshifts of $\sim$18000 galaxies were measured with a median error of $50~{\rm km~s^{-1}}$.
The 288 spectroscopic galaxies are selected as the member galaxies \citep{2017A&A...599A..81M}. 
By using the velocity dispersion $\sigma~=~760~\pm ~28~{\rm km~s^{-1}}$ estimated by using the member galaxies \citep{2017A&A...599A..81M} and 
the bolometric X-ray luminosity $L_{X}~=~4.5~\times~10^{43}~{\rm  erg~s^{-1}}$, we compared with the $\sigma-L_{X}$ 
relationship of the X-ray selected clusters. From the relationship of table 3 in \citet{2011A&A...526A.105Z}, 
the expected bolometric luminosity is $\sim2\times10^{44}~{\rm erg~s^{-1}}$, 
and it is found that the bolometric luminosity is relatively lower at a given velocity dispersion. 

On the other hand, the temperature and velocity dispersion are in good agreement 
with the $T-\sigma$ scaling relation within a factor of $\sim$1.5 \citep{2000ARA&A..38..289M}. 
These features may show that all derived quantities are roughly consistent 
with the cluster scaling relations except for the low-luminosity and high-entropy features. 
It appears that the low central gas density results in the low luminosity and high entropy.

%

We compared the spatial distribution of the X-ray emission with the member galaxy spatial distribution.
Figure \ref{fig:img} shows that the X-ray emission in the south is much fainter,
while the number of the member galaxies in the south, 67 galaxies, 
is not significantly smaller than those of the other regions (east: 67, north: 54, west: 98).
There is no significant discrepancy between radial profiles of the number density of the member galaxies in all directions.  
Thus, X-ray surface brightness distribution spatially differs from the galaxy distribution.

\subsection{Possible scenario}
The A1631 shows the following observational features; 
(1) higher entropy, $K \gtrsim 400~{\rm keV~cm}^2$, at the central region ($r<0.1~r_{200}$),
(2) the relatively lower bolometric luminosity for the given velocity dispersion,
(3) spatial difference between the X-ray surface brightness distribution and the galaxy distribution. 

One possible explanation for these phenomena is that the cluster is a post-merger system and 
went through the counterpart as the Bullet cluster system \citep{2004ApJ...606..819M}. 
The merger could have dispersed the low-entropy gas in the center and mixed it with the high-entropy ICM at 
larger radii, which leads to the central high entropy and the flatter entropy profile \citep{2011ApJ...728...54Z}.
The ICM is stripped by ram pressure and the X-ray luminosity is expected to be low, relative to that predicted from the velocity dispersion.
The spatial difference between the X-ray surface brightness and the galaxy distribution 
is also explained by this scenario as seen in the Bullet cluster system, even though, 
we can not find obvious evidence for merger activities, such as density jumps, 
which may be due to the large size of Suzaku's PSF. Future observations with higher angular resolution 
may be able to see the interaction effects.

This cluster is located in the Shapley Supercluster ($z~\sim~0.05$), which is the most massive known structure 
in the local Universe \citep{2014MNRAS.445.4073C} and which may provide a particularly rich environment for merger activity. 
Assuming the typical collision velocity of 600 km $^{-1}$ expected from the concordance 
$\Lambda$ cold dark matter model \citep{2012MNRAS.419.3560T} and the half of age of the Universe, 
no clusters are found around $z~\sim~0.05$ within the region in the NED web site site\footnote{https://ned.ipac.caltech.edu} as candidates of the partner. 
If the post-merger scenario is correct, the subcluster lies in A1631. 
Future weak lensing observations may identify the dark matter clumps originated in the merging system.


\section{Summary}
We have analyzed the X-ray data of the LSB cluster A1631, which has the lowest surface brightness 
in a cluster sample of the ROSAT All-Sky Survey. 
Based on the Suzaku observation data, the observed X-ray morphology has no strong peak and is irregular. 
We have examined the spatial distribution of the physical properties, gas temperature, density, and entropy, 
from the Suzaku and XMM-Newton data, out to $\sim 1.5~{\rm Mpc}$. The gas density is comparatively 
low (${\rm a~few} \times~10^{-4}~ {\rm cm^{-3}}$) 
for the observed high temperature ($\sim3 ~{\rm keV}$) compared with those of X-ray-selected clusters. 
The central gas entropy ($\gtrsim400~{\rm keV~cm^2}$) is the highest among the known nearby clusters. 
The bolometric luminosity is approximately 1/3 of that expected from 
the observed temperature-luminosity relationships. 
These properties are also observed in the low surface brightness cluster A76. 
By using the rich optical data, we have compared the X-ray properties with the optical properties.
We have found that the spatial distributions of the member galaxies and the X-ray emission differ 
and that the bolometric luminosity is relatively lower at the given velocity dispersion.
We have proposed the post-merger scenario to explain these observational properties.
\section*{Acknowledgement}
We thank all the Suzaku team members for their support. 

YB is grateful to Professor Y. Tawara (Nagoya University) for giving various 
useful advice in the analysis and thanks Professor Y. Fujita (Osaka University) 
for his insightful comments in the discussion. 
This work was supported in part by JSPS KAKENHI grant number 16K05295 (NO). 
HB and GC acknowledge support from the DFG Transregio Program TR33 and
the Munich Excellence Cluster “Structure and Evolution of the Universe”
and also thank Kavli IPMU for the hospitality and the support during
our stay as well as the Excellence Cluster.
We thank Takaya Ohashi for the workshop, which helped this collaboration.
GWP acknowledges funding from the European Research Council under the European
 Union’s Seventh Framework Programme (FP7/2007−2013)/ERC grant agreement No. 340519.

The authors are grateful to the anonymous referee for his comprehensive comments 
and useful suggestions, which improved the paper very much.

\end{document}